\documentclass{appolb}
\usepackage{graphicx}
\usepackage{slashed}
\usepackage{amsmath, amssymb, graphics, setspace}
\usepackage{epstopdf}
\usepackage{multirow}
\usepackage{tikz-feynman}
\usepackage{subcaption}
\newcommand{\bvec}[1]{\mbox{\boldmath ${#1}$}}

\begin{document}
\title{Some Phenomenological Aspects of Kaon Photoproduction in the Extreme Kinematics}%
\author{S.Sakinah, S.Clymton and T. Mart
\address{Departemen Fisika, FMIPA, Universitas Indonesia, Depok 16424, Indonesia }
\\}
\maketitle
\begin{abstract}
We have investigated phenomenological aspects of kaon
photoproduction in three different extreme kinematics.
The first kinematics of interest is the threshold region.
At the threshold we have investigated the convergence of kaon 
photoproduction amplitude by expanding the square of the 
amplitude in terms of the ratio $m_K/m$, where $m_K$ is 
the mass of kaon and $m$ is the averaged mass of nucleon 
and $\Lambda$-hyperon. The amplitude is calculated from 
the appropriate Feynman diagrams by using the pseudovector 
theory. The contact diagram as a consequence of the PCAC 
hypothesis is also taken into account in the amplitude. 
Our finding indicates that the convergence can be only 
achieved if the amplitude was expanded up to at least 
$12^{\rm th}$ order. As a consequence, applications of 
some theoretical calculations based on the expansion 
the scattering amplitude, such as the Low Energy Theorem 
or Soft Kaon Approximation, cannot be easily managed in 
kaon photoproduction. The second kinematics is the forward
region, where we could assume only $t$-channel contributes to the 
process. Here we have investigated the effect of amplitude expansion
on the extraction of the coupling constants
$g_{K^+\Lambda p}$ and $g^V_{K_1^+\Lambda p}$.
The last kinematics is the backward region, where we have also
assumed that only $u$-channel survives and we could extract 
the leading coupling constants $g_{K^+\Lambda p} $ and 
$g_{K^+\Sigma^0 p}$.
\end{abstract}
  
\newpage
\section{Introduction}
In the previous works we have studied kaon photo- and electroproduction 
near their production thresholds by using the pseudo-scalar (PS) and 
pseudo-vector (PV) theories \cite{Mart10,Mart:2011ez}. Extended studies 
to cover not only the threshold energy but also the resonance region have 
been also performed \cite{Mart:2015jof,Clymton:2017nvp,Mart:2017mwj}.
These studies indicate that  compared to the PV coupling the PS one leads 
to a better agreement with experimental data. This is the reason that 
the PS coupling is commonly used in the phenomenological investigations 
of kaon photoproduction. We note, however, that this is not the case in 
pion production, where the PV coupling can be and mostly used in the 
production formalism. The reason is that the cross section of kaon 
production obtained in experiments is two order of magnitude smaller 
than that of pion production, whereas the Feynman diagrams and the 
coupling constants used in the theoretical formulation of both 
processes are similar. Since in photoproductions the pion thresholds 
are the lowest, pion photoproduction has been used as the main reaction 
for investigating the Low Energy Theorem (LET) \cite{Vainzakh,Gaffney}. 
LET is derived by using the PV coupling and expanding the pion 
photoproduction amplitude in terms of the ratio of pion and nucleon 
masses. Since this ratio is considerably small and the expansion 
can quickly converge, terms with higher orders can be obviously 
neglected. We also note that the successful Chiral Perturbation 
Theory formulation assumes that pion is a pseudovector particle.

On the other hand, kaon mass is much heavier than pion mass. Therefore, for kaon photoproduction LET has not been seriously considered since the convergence cannot be easily achieved. However, an attempt to derive LET for the radiative decay width of charged kaon $K \rightarrow l+\nu+\gamma $ within the so-called Soft Kaon Approximation (SKA) was performed more than five decades ago \cite{SKA_Ronald}. Comparison with the result obtained from the Pole Dominance Approximation \cite{pda} reveals the fact that the result of the SKA is shifted upward by approximately 20\%. Such a significant difference indicates that further studies of SKA are strongly required. Presumably, the problem originates from the intrinsic properties of kaon as compared to the pion, e.g., the heavier mass of kaon. A systematic and careful investigation of the convergence of kaon scattering amplitude is therefore very important to this end. As a first step, we might investigate the convergence of kaon photoproduction amplitude, for which experimental data are abundant nowadays.

Furthermore, kaon photoproduction at forward and backward angles is also
of interest, since at the two extreme directions contributions of $t$ and 
$u$ channels, would be dominant. As a consequence, theoretical
formulation of kaon photoproduction in both cases could be extremely 
simplified. This is valid not only in kaon photoproduction, but also in
the electromagnetic production of meson in general.
For instance, in Ref.~\cite{Mart:2008sw} it is shown that it is
possible to extract the pion electromagnetic form-factor 
by merely using the $t$-channel diagram, provided that 
only the pion electroproduction data obtained in the forward direction 
are used in the analysis. 
It is also important to mention here that forward angles are 
very decisive for hypernuclear production, since only in this 
kinematics the nuclear cross section is sufficiently large. The nuclear
form factor suppresses this cross section strongly as the
kaon scattering angle increases. Since the elementary operator 
for this purpose is constructed from the kaon photoproduction 
amplitude, an accurate description for the forward angles 
production is inevitable \cite{appl_nuclei}.
To check whether or
not the $u$-channel contribution dominates the kaon photoproduction
amplitude, measurement of kaon photoproduction at backward angles 
has been also performed at SPring8
more than a decade ago \cite{Hicks:2007zz}.

In this paper we investigate the convergence of kaon photoproduction amplitude by expanding the squared amplitude, which is proportional to the cross section, in terms of the ratio $m_K/m$, where $m_K$ is the mass of kaon and $m$ is the averaged mass of nucleon and $\Lambda$-hyperon. For this purpose, we construct the minimal model that can explain experimental data very close to the threshold. We use the suitable Feynman diagrams with PV coupling based on our previous studies \cite{Mart10,Mart:2011ez}. The analytic calculation of the amplitude expansion was performed with the help of Mathematica software. The same method is also used to derive the reaction amplitudes in the forward and backward angles.

This paper is organized as follows. In Section \ref{Sect:2} we present 
the kinematics of the photoproduction process. 
Section \ref{sec:expansion} discusses the expansion of the photoproduction 
amplitude in general. The corresponding numerical result is given in Section 
\ref{sec:result}. Section \ref{sec:forward} is devoted for the discussion
of kaon photoproduction in the forward region. The case of backward angles
photoproduction is given in Section \ref{sec:backward}.
We will summarize our finding in Section \ref{sec:summary}. 
The expansion of the photoproduction amplitude up to $9^{\rm th}$ order 
is given in Appendix \ref{append:Amplitude}.

\section{Kinematics}\label{Sect:2}
Let us consider the kaon photoproduction process 
\begin{eqnarray}
  \gamma (k) + p(p_p) \longrightarrow K^{+}(q) + \Lambda (p_\Lambda) ,
  \label{eq:reaksi}
\end{eqnarray}
where the four-momenta of the photon $\gamma$, proton $p$, kaon 
$K^+$ and hyperon $\Lambda$ are 
explicitly
indicated.
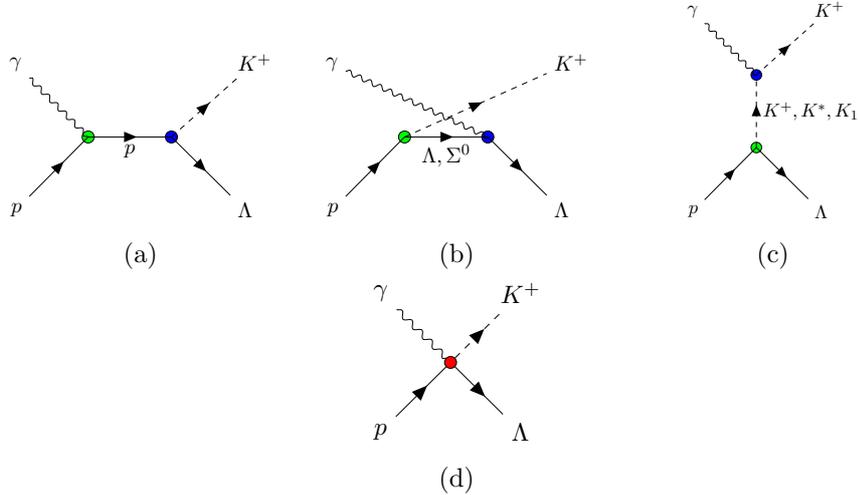
\begin{figure}
  \centering
  \begin{subfigure}[b]{0.32\textwidth}
  \centering
\resizebox{0.9\columnwidth}{!}{%
\begin{tikzpicture}
 \begin{feynman}
    \node[circle,scale=0.6,draw=black,fill=green!100] (b);
    \node[circle,scale=0.6,draw=black,fill=blue!100,right=of b] (c);
    \vertex (b);
    \vertex [right=of b] (c);
    \vertex [above left=of b] (f1) {\(\gamma\)};
    \vertex [below left=of b] (a) {\(p\)};
    \vertex [above right=of c] (f2) {\(K^+ \)};
    \vertex [below right=of c] (f3) {\(\Lambda\)};
 
    \diagram* {
      (a) -- [fermion] (b) -- [photon] (f1),
      (b) -- [fermion, edge label'=\(p\)] (c),
      (c) -- [charged scalar] (f2),
      (c) -- [fermion] (f3),
    };
\end{feynman}
\end{tikzpicture}%
}
    \caption{}
  \end{subfigure}
  \begin{subfigure}[b]{0.32\textwidth}
  \centering
\resizebox{0.9\columnwidth}{!}{%
\begin{tikzpicture}
\begin{feynman}
    \node[circle,scale=0.6,draw=black,fill=green!100] (b);
    \node[circle,scale=0.6,draw=black,fill=blue!100,right=of b] (c);
    \vertex (b);
    \vertex [right=of b] (c);
    \vertex [above left=of b] (f1) {\(\gamma\)};
    \vertex [below left=of b] (a) {\(p\)};
    \vertex [above right=of c] (f2) {\(K^+\)};
    \vertex [below right=of c] (f3) {\(\Lambda\)};
 
    \diagram* {
      (a) -- [fermion] (b) -- [charged scalar] (f2),
      (b) -- [fermion, edge label'=\(\Lambda{, }\, \Sigma^0\)] (c),
      (f1) -- [photon] (c),
      (c) -- [fermion] (f3),
    };
\end{feynman}
\end{tikzpicture}%
}
    \caption{}
  \end{subfigure}
  \begin{subfigure}[b]{0.32\textwidth}
  \centering
\resizebox{0.6\columnwidth}{!}{%
\begin{tikzpicture}
\begin{feynman}
    \node[circle,scale=0.6,draw=black,fill=green!100] (b);
    \node[circle,scale=0.6,draw=black,fill=blue!100,above=of b] (c);
    \vertex (b);
    \vertex [above=of b] (c);
    \vertex [above left=of c] (f1) {\(\gamma\)};
    \vertex [below left=of b] (a) {\(p\)};
    \vertex [above right=of c] (f2) {\(K^+\)};
    \vertex [below right=of b] (f3) {\(\Lambda\)};
 
    \diagram* {
      (a) -- [fermion] (b) -- [fermion] (f3),
      (b) -- [charged scalar, edge label'=\(K^+{,}\, K^*{,} \,K_1\)] (c),
      (f1) -- [photon] (c),
      (c) -- [charged scalar] (f2),
    };
\end{feynman}
\end{tikzpicture}%
}
    \caption{}
  \end{subfigure}
  \begin{subfigure}[b]{0.32\textwidth}
  \centering
\resizebox{0.6\columnwidth}{!}{%
\begin{tikzpicture}
\begin{feynman}
    \node[circle,scale=0.5,draw=black,fill=red!100] (b);
    \vertex [above left=of b] (f1) {\(\gamma\)};
    \vertex [below left=of b] (a) {\(p\)};
    \vertex [above right=of b] (f2) {\(K^+\)};
    \vertex [below right=of b] (f3) {\(\Lambda \)};
 
    \diagram* {
      (a) -- [fermion] (b) -- [fermion] (f3),
      (f1) -- [photon] (b),
      (b) -- [charged scalar] (f2),
    };
\end{feynman}
\end{tikzpicture}%
}
    \caption{}
  \end{subfigure}
  \caption{Feynman diagrams for the Born terms of the $K^+\Lambda$ photoproduction. The corresponding
  $s$, $u$, and $t$ channels are shown by the diagrams (a), (b), and (c), respectively, whereas
  the contact (seagull) term is given by diagram (d).
  \label{fig:feynman}}
\end{figure}

To derive the photoproduction amplitude we use the first order Feynman diagrams shown in Fig. \ref{fig:feynman}. Following the previous works on pion photoproduction, here we consider kaon as a pseudovector particle, instead of pseudoscalar one. As will be explained in the next Section, the choice is also supported by the fact that the expansion of the PS amplitude has a serious problem since the PS amplitude contains the large ${\cal O}(m/m_K)$ terms, instead of the small and expandable ${\cal O}(m_K/m)$ terms. The conservation of axial-vector current requires the kaon to be massless. Therefore, at $m_K\rightarrow 0$ and $q=0$ (at threshold), the amplitude to the first order of electromagnetic coupling can be written as
\begin{equation}\label{eq:constraint}
\mathcal{M}({m_K\to 0,q=0}) = \bar{u}_\Lambda(p_\Lambda)\left[-\frac{ g_{K^+\Lambda p}}{ m_p+m_\Lambda}\,  \gamma_5 \slashed{\epsilon}\right]u_p(p_p) ~.
\end{equation}
This term is known as the contact (seagull) term, which contributes in 
the zeroth order of amplitude and it is considered as a  consequence of 
the partially conserved axial current (PCAC) hypothesis~\cite{Adler139}. 
Besides the Born and contact terms we also include two more intermediate
states that have been proven to be important to achieve a better agreement 
with experimental data. They are the $K^*$ and $K_1$ vector meson 
resonances. The total amplitude reads 

\begin{eqnarray}\label{eq:amplitude}
\mathcal{M} &=& \bar{u}_\Lambda(p_\Lambda)\left[i \frac{g_{K^+\Lambda p}}{m_\Lambda+m_p }\gamma_5\slashed{q} \frac{\slashed{p}_p+\slashed{k}+m_p}{s-m_p^2}(e\slashed{\epsilon}+i\sigma^{\mu\nu}\epsilon_\mu k_\nu \mu_p)\right.\nonumber\\
&&\left.+ i\sigma^{\mu\nu}\epsilon_\mu k_\nu \mu_\Lambda \frac{\slashed{p}_\Lambda-\slashed{k}+m_\Lambda}{u-m_\Lambda^2}\frac{i g_{K^+\Lambda p}}{m_\Lambda+m_p }\gamma_5\slashed{q} \right.\nonumber \\
&& \left.+i e\frac{g_{K^+\Lambda p}}{m_\Lambda+m_p }\gamma_5(\slashed{q}-\slashed{k})\frac{2q\cdot\epsilon}{t-m_k^2}-\frac{ g_{K^+\Lambda p}}{m_p+m_\Lambda}\gamma_5\slashed{\epsilon}\right. \nonumber \\
&& \left.+\frac{i}{M(t-m^2_{K^*}+im_{K^*}\Gamma_{K^*})}\left\{g^V_{K^*\Lambda p}\gamma_\mu-\frac{g^T_{K^*\Lambda p}}{m_p+m_\Lambda}i\sigma^{\mu\nu}(q_K-k)_\nu \right\}\right.\nonumber\\
&& \left.\times i\varepsilon_{\mu\nu\rho\sigma} \epsilon^\nu k^\rho q^\sigma_K g_{K^*K^+\gamma}+\frac{i}{M(t-m^2_{K_1}+i m_{K_1}\Gamma_{K_1})}\Bigl\{ g^V_{K_1\Lambda p}\gamma^\mu\gamma_5\right.\nonumber\\
&& \left.+ \frac{g^T_{K_1\Lambda p}}{m_p+m_\Lambda}(\slashed{p}_\Lambda-\slashed{p}_p)\gamma^\mu \gamma_5\Bigr\}(q_K\cdot\epsilon k_\mu-q_K\cdot k \epsilon_\mu)g_{K_1 K^+\gamma}\right.\nonumber\\
&& \left.+ i \sigma^{\mu \nu}\epsilon_\mu k_\nu \mu_T \frac{\slashed{p}_\Sigma-\slashed{k}+m_\Sigma}{u-m_\Sigma^2}\frac{i g_{K^+\Sigma^0 p}}{m_\Sigma+m_p}\gamma_5\slashed{q}\right]u_p(p_p) ~,
\end{eqnarray}
where $\epsilon$ is the photon polarization,  $\mu_p$ and $\mu_\Lambda$  
represent the magnetic moments of proton and $\Lambda$-hyperon, respectively, 
while $s$, $t$ and $u$ indicate the Mandelstam variables. In 
Eq.~(\ref{eq:amplitude}) we have introduced $M=1$ GeV to make the 
transition strengths $g_{K^*K^+\gamma}$ and $g_{K_1K^+\gamma}$ dimensionless.
Note that since in the phenomenological studies of kaon photoproduction 
the two transition strengths cannot be explicitly separated, we define
$G^{V,T}_{K^*} = g_{K^* K^+\gamma}\, g_{K^* \Lambda p}^{V,T}$ and 
$G^{V,T}_{K_1} = g_{K_1 K^+\gamma}\, g_{K_1 \Lambda p}^{V,T}$.

For the calculation of cross section with the 
full amplitude it is customary to decompose
the transition amplitude in Eq.~(\ref{eq:amplitude}) into
the gauge and Lorentz invariant matrices $M_{j}$, i.e., \cite{Lee:1999kd}
\begin{eqnarray}
  \label{eq:mfi}
\mathcal{M} &=& \bar{u}(p_\Lambda) \sum_{j=1}^4 A_j(s,t,u) M_j u(p_p) ~,
\end{eqnarray}
where
\begin{eqnarray}
M_{1} &=& \gamma_5\slashed{\epsilon}\slashed{k} ~,\\
M_{2} &=& 2\gamma_5\left( q\cdot\epsilon P\cdot k-q\cdot k~P \cdot\epsilon\right) ~,\\
M_{3} &=& \gamma_5(q\cdot k\slashed{\epsilon}-q\cdot\epsilon\slashed{k}) ~,\\
M_{4} &=& i\epsilon_{\mu\nu\rho\sigma}\gamma^\mu q^\nu \epsilon^\rho k^\sigma ~,
\end{eqnarray}
with $P=(p_p+p_\Lambda)/2$ and $\epsilon_{\mu\nu\rho\delta}$ the Levi-Civita tensor. 
The form functions $A_i$ given in Eq.~(\ref{eq:mfi})
can be used to calculate the cross section.

The transition amplitude ${\cal M}$ in Eq.~(\ref{eq:amplitude}) 
can be also written within the PS 
theory by replacing ${g_{K^+\Lambda p}}\gamma_5\slashed{q}/(m_\Lambda+m_p)$ 
in the $s$ and $u$ channels and  
${g_{K^+\Lambda p}}\gamma_5(\slashed{q}-\slashed{k})/(m_\Lambda+m_p)$ 
in the $t$ channel with $g_{K^+\Lambda p}\gamma_5$ \cite{Mart10}.
However, it is found that the expansion of the PS amplitude
is difficult because the expansion contains the terms that are proportional
to $x^{-1}$, where $x=m_K/m$. Thus, to achieve the convergence 
is a daunting task in this case. The problem originates from the 
fact that in the PS theory the amplitude is not proportional 
to the kaon momentum $q$, as in the case of PV theory
[see Eq.~(\ref{eq:amplitude})].

\section{Expansion of the Amplitude}
\label{sec:expansion}
The transition amplitude of kaon photoproduction obtained from the Feynman diagrams shown in Fig. \ref{fig:feynman} is given in Eq.~(\ref{eq:amplitude}). The form functions $A_i$ are intended for the calculation of the cross section in the data fitting as well as for obtaining the result of full calculation which will be compared with the approximations made in the present work~\cite{Mart10}.  Since we limit our study to energies very close to threshold, we do not use nucleon resonances in our calculation. The latter is also important to significantly simplify the amplitude formulation.

\begin{table}[t]
  \centering
\renewcommand{\arraystretch}{1.1}
  \caption{Coupling constants extracted from fittings to experimental
    data in the previous \cite{Mart10} and present works and the
    corresponding $\chi^2/N$.}
  \label{tab:Couplingconstants}
  \begin{tabular}[c]{lcc}
    \hline\hline\\[-2ex]
   Coupling constant  & Previous & Present \\ [0.5ex]
   \hline\\[-2.2ex]
   $g_{K^+\Lambda p}/\sqrt{4\pi}$  &$-3.80$& $-3.00$  \\
   $g_{K^+\Sigma^0 p}/\sqrt{4\pi}$ &$~~1.20$&$~~1.30$  \\
   $G^V_{K^*}/4\pi$                &$-0.79$& $-0.73$  \\
   $G^T_{K^*}/4\pi$                &$-0.04$& $~~0.70$ \\
   $G^V_{K_1}/4\pi$                &$~~1.19$& $~~0.80$ \\
   $G^T_{K_1}/4\pi$                &$-0.68$& $-1.50$  \\[0.5ex]
   $\chi^2/N$                      &$1.526$ &\, $1.739$ \\[0.5ex]
   \hline\hline
  \end{tabular}
\end{table} 

There are very few experimental data points near the threshold, 
as will be shown in the next Section. The closest ones to the
threshold are obtained from the SAPHIR data \cite{Glander:2003jw}.
Since we exclude the nucleon resonances the calculated full amplitude
in the present calculation differs from our previous model \cite{Mart10}. As a 
consequence, we must refit the coupling constants in order to
obtain a reasonable amplitude. The result
is shown in Table \ref{tab:Couplingconstants}, where we compare
our present result with that of the previous PV calculation \cite{Mart10}.
From Table \ref{tab:Couplingconstants} it appears that the coupling
constants obtained from refitting the experimental data differ, 
although not dramatically, from those of previous work. The larger value of
$\chi^2/N$ obtained in the present work is understandable because the number of
free parameters decreases due to the exclusion of nucleon and hyperon 
resonances.

The square of photoproduction amplitude is expanded in terms of the ratio between kaon and baryon masses in the reaction. Different from the pion photoproduction, in the kaon photoproduction the baryon in the initial and final states are proton and $\Lambda$-hyperon, respectively. Therefore, we define the baryon mass $m$  as the average of the proton mass $m_p$ and $\Lambda$-hyperon mass  $m_\Lambda$ and the ratio becomes $x=m_K/m=2m_K/(m_p+m_\Lambda)$. Furthermore, to simplify the formalism, we also make the assumption that $m_p\approx m_\Lambda$ in the present work.

In this work the expansion of the square of amplitude has been performed up to $n=20$, in order to reach the convergence. The expansion has been done analytically with the help from the Mathematica software. We calculate the expansion of the squared amplitude instead of the amplitude itself because we find it is more simple and practical to use and to compare with experimental data as well as the result of other calculations. We aware that there exist a number of studies which compare the results with multipoles amplitudes. Nevertheless, for this exploratory study we limit our calculation to the total cross section. At the threshold the (reduced) total cross section of kaon photoproduction can be written as 
\begin{eqnarray}
\frac{|\bvec{k}|}{|\bvec{q}|} \sigma_{\rm tot}&=&\frac{1}{64 \pi s} \mathcal{|M|}^2
\end{eqnarray}
with 
\begin{eqnarray}
\mathcal{|M|}^2 &=& \sum_{n=0}^{20} \mathcal{|M|}_{(n)}^2 ~=~ 
\mathcal{|M|}_{(0)}^2+\mathcal{|M|}_{(1)}^2+\mathcal{|M|}_{(2)}^2+\cdots
\end{eqnarray}
where the subscript indicates the order of expansion. Therefore, $\mathcal{|M|}_{(n)}^2\propto x^n$. The formulas of $\mathcal{|M|}_{(n)}^2$ with $n$ up to 9 are given in Appendix \ref{append:Amplitude}. For higher order amplitudes the formulas are very complicated and too long to be written in this paper, although it is produced by Mathematica and relatively simple for numerical computation. Nevertheless, for the sake of completeness in the present work we still calculate the cross section with $n$ up to 20, in spite of the fact that to reach the convergence the number of order is significantly less than 20, as will be discussed in the next Section.

\section{Numerical Results}
\label{sec:result}
\begin{figure}[t]
\centerline{%
\includegraphics[width=13cm]{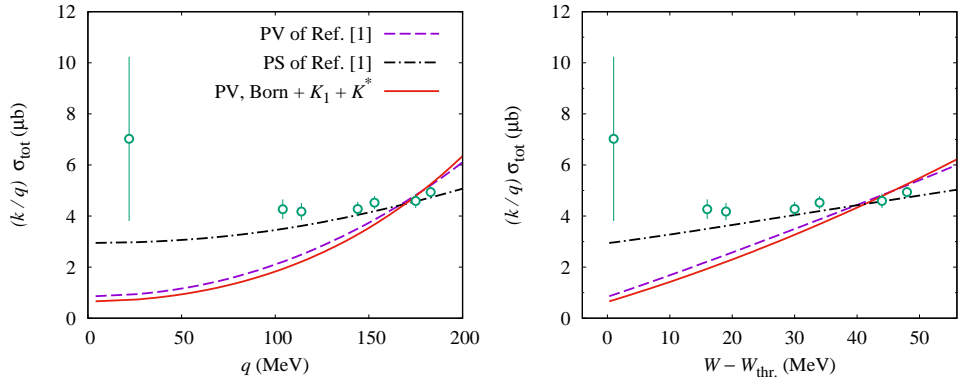}}
\caption{Reduced total cross section as a function of the kaon three-momentum 
  $q=|\bvec{q}|$ (left panel) and the total c.m. energy above the threshold
  energy (right panel) 
  obtained by using the PV (dashed lines) and
  PS (dot-dashed lines) couplings, where the Born, $K^*$, $K_1$, hyperon, 
  and nucleon resonance terms are included \cite{Mart10}. 
  Solid lines are obtained from 
  the calculation that includes the Born, $K^*$, and $K_1$ terms only.
  Open circles indicate the CLAS 2006 data \cite{CLAS}.}
\label{Fig:PVmodel}
\end{figure}

The momentum and energy distributions of the calculated reduced total cross section are shown in Fig. \ref{Fig:PVmodel}, where we also compare the result of previous calculations obtained by using PS and PV couplings \cite{Mart10} along with the available experimental data \cite{CLAS}. As discussed above as well as in Ref.~\cite{Mart10} the PS model yields a better agreement with experimental data. This is clearly shown in Fig. \ref{Fig:PVmodel}. Furthermore, we can also see that the PV model without hyperon and nucleon resonances produces a slightly different cross section compared to the full calculation. This result indicates that without hyperon and nucleon resonances the PV model can still nicely work and provide a good framework for the purpose of comparison with the present work. Furthermore, from Fig. \ref{Fig:PVmodel} we find that the calculated reduced total cross section at threshold is around 0.65 $\mu$b. This is clearly smaller than the experimental data data near the threshold. However, we also observe that the experimental error bar in this case is significantly large. Thus, we still believe that for the sake of comparison made in the present work Fig. \ref{Fig:PVmodel} justifies that we can use the simplified model, i.e., the model constructed from Born terms along with the $K^*$ and $K_1$ vector mesons.

\begin{table}[t]
\renewcommand{\arraystretch}{1.1}
  \centering
  \caption{Properties of the particles considered in this study \cite{PDG}.}
  \label{tab:PDG}
\begin{tabular}{ccrcccc}
\hline\hline\\[-2ex]
 Particle & $S$ & $J^{P}$ & $~~I~$ & $\mu$ (n.m.) & Mass & Width\\
 &&&&& (MeV)& (MeV)\\
 [0.5ex]
 \hline\\[-2.2ex]
 $p$&0&$\frac{1}{2}^{+}$&$\frac{1}{2}$&$2.79284734$&$938.2721$&-\\
 $n$&0&$\frac{1}{2}^{+}$&$\frac{1}{2}$&$-1.91304273$&$939.5654$&-\\
 $K^{+}$&$ 1$&$0^{-}$&$\frac{1}{2}$&-&$493.677$~~&-\\
 $\Sigma^{0}$&$-1~~$&$\frac{1}{2}^{+}$&$1$&$1.61\pm 0.08$ &$ 1192.642$&-\\
 $\Lambda$&$-1~~$&$\frac{1}{2}^{+}$&$0$&$-0.613\pm 0.004$&$1115.683$&-\\
 $K^{*+}$&$1$&$1^{-}$&$\frac{1}{2}$&-&$891.76\pm 0.25$&$50.3\pm0.8$\\
 $K_1$&1&$1^{+}$&$\frac{1}{2}$&-&$1272\pm 7$&$90\pm 20$\\[0.5ex]
 \hline\hline
\end{tabular}
\end{table}

\begin{table}[t]
  \centering
  \caption{Contribution of the $n$-th order photoproduction 
    amplitude to the cross section at threshold. }
  \label{tab:Result}
  \begin{tabular}[c]{cc}
  \hline\hline\\[-2ex]
   Order of the~~    & Cross Section \\
   expansion~~& ($\mu$b) \\ \hline\\[-2ex]
   $0^{\rm th}$& $~33.98$\\
   $1^{\rm st}$ & $~19.38$\\
   $2^{\rm nd}$ & $-37.89~$ \\
   $3^{\rm rd}$ & $-10.58$\\
   $4^{\rm th}$ & $~12.56$\\
   $5^{\rm th}$ & $-2.96$\\
   $6^{\rm th}$ & $-0.31$\\
   $7^{\rm th}$ & $~~1.78$\\
   $8^{\rm th}$ & $~~0.03$\\
   $9^{\rm th}$ & $~~0.53$\\
   $10^{\rm th}$& $~~0.66$\\ 
   $11^{\rm th}$& $~~0.47$\\
   $12^{\rm th}$ & $~~0.54$ \\
   $13^{\rm th}$ & $~~0.55$\\
   $14^{\rm th}$ & $~~0.53$\\
   $15^{\rm th}$ & $~~0.53$\\
   $\cdots$ & $\cdots$\\
   $20^{\rm th}$& $~~0.53$\\
   \hline\hline
  \end{tabular}
\end{table}

\begin{figure}[t]
\centerline{%
\includegraphics[width=11.5cm]{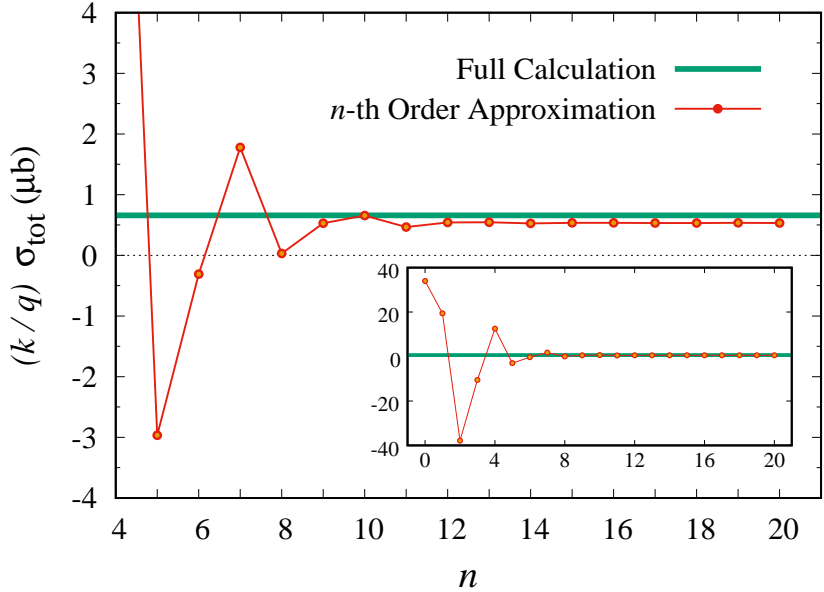}}
\caption{Convergence of the reduced total cross section as a function of
  the number of order of the expansion ($n$) with $n > 4$.  
  The red line indicates the
  approximated total cross section, whereas the green line shows the
  result of full calculation. The inset displays the similar comparison 
  but started with $n\geq 0$.} 
\label{Fig:order}
\end{figure}

By using the PV model with parameters given in the third column of Table~\ref{tab:Couplingconstants} and the particle properties taken from the Particle Data Group (PDG) listed in Table~\ref{tab:PDG} we calculate the reduced total cross section at threshold for each of the expansion order. The result is listed in Table~\ref{tab:Result} and graphically displayed as a function of $n$ (order of expansion) in Fig.~\ref{Fig:order}, where the bold green line indicates the full order calculation as our reference value. From Table \ref{tab:Result} and Fig.~\ref{Fig:order} it is apparent that the lowest order calculation yields very large discrepancy with the full calculation. Figure~\ref{Fig:order} shows that the expansion starts to converge at the $9^{\rm th}$ order. Although at this order a small discrepancy still exists (about 6\%), we have also to admit that we have made an assumption for baryonic mass $m$, which could also create sizable error. 

The result discussed above indicates that compared to the case of pion photoproduction the LET for kaon photoproduction would require higher order terms. Actually, by excluding the $K^*$ and $K_1$ contributions  we can achieve the convergence relatively faster, i.e., it converges at the $n=7$ order. However, it is well known that the Born terms alone cannot reproduce the kaon photoproduction data, even near the threshold. This is in contrast to the pion photoproduction case. Note that in this work we expand the square of amplitude, instead of the amplitude itself. As a consequence, it is natural if the order of expansion required in the first case is twice larger than that of the second case. Nevertheless, as we explained above, the convergence of this squared amplitude is reached with $n=9$. Therefore, the convergence of linear amplitude would be reached with $n > 4 $. This is still not comparable to the case of pion photoproduction.

The expansion of squared amplitude given in Appendix  \ref{append:Amplitude} reveals the fact that there are zeroth order and first order squared amplitudes that depend solely on the coupling constant $g_{K^+\Lambda p}$ [see Eqs.~(\ref{eq:zeroth-order}) and (\ref{eq:first-order})]. Should we propose a LET by using only these term, as in the case of pion photoproduction, then we would find that the value of $g_{K^+\Lambda p}$ must be much smaller than the prediction of SU(3) symmetry if the calculated cross section must be comparable to experimental data. The reason is that the cross section of kaon photoproduction is much smaller than that of the pion one. To sum up, we can safely say that the theories which are based on low order expansion of kaon mass, such as SKA and Chiral Perturbation Theory, should include higher order terms in order to produce the experimental data. 

Should we try to create a LET by using higher order terms in our formalism, then the main problem would be the number of unknown parameters. We can also see that the expansion described above reduces the number of unknown parameters, i.e., the coupling constants of $K_1$ and $K^*$ intermediate states. From Eq.~(\ref{eq:amplitude}) we see that there are two coupling constants for each of them. From our assumption to generalize the baryonic masses, i.e., $m_p\equiv m_\Lambda \equiv m$, the two couplings reduces to only one coupling [see Eqs.~(\ref{eq:GKS_coupling}) and (\ref{eq:GK1_coupling})]. As a result, in this formalism we have only four unknown coupling constants. Nevertheless, it is by no means easy to derive LET from these 4 parameters. We have to find other mechanisms to reduce them. This is the interest of our future study, i.e., to create higher order LET for kaon photoproduction. Furthermore, in the future we will consider the expansion of amplitude, instead of the squared one described in the present work.

\section{Kaon Photoproduction in the Forward Direction}
\label{sec:forward}

In the forward direction it is well known that the denominator of the 
$t$-channel propagator  ($t-m_K^2$) is small. As a consequence, the 
corresponding propagator becomes large and the $t$-channel contribution 
dominates the production amplitude. The total amplitude given by 
Eq.~(\ref{eq:amplitude}) can be reduced and the unknown parameters 
can be limited to the coupling constants $g_{K^+\Lambda p}$ and 
$g^V_{K_1^+\Lambda p}$ only. Since very close to the  threshold 
there exist no experimental data in the forward angles, we extend
the model to cover the resonance region, where $|\bvec{q}|\neq 0$. 

It is important to note that the $t$-channel exchange is not 
individually gauge invariant. In the PV coupling theory, the nucleon 
exchange in the $s$-channel and the contact term must be included to 
restore the gauge invariance. In the PS coupling the electric part 
of the $s$-channel alone is sufficient to restore the gauge invariance.
This is in contrast to the anomalous magnetic moment part, which is 
already gauge invariant by design, but not included in this 
work \cite{Guidal}.

\begin{table}[t]
\centering
  \caption{Coupling constants $g_{K^+\Lambda p}$ and $g^V_{K_1^+\Lambda p}$
    extracted from the expanded photoproduction amplitude up to
     n-th order in the forward region. }
  \label{tab:forward_result}
  \begin{tabular}[c]{ccccccc}
  \hline\hline\\[-2ex]
   {Order}    & \multicolumn{2}{c}{$K^+$ exchange}& & \multicolumn{3}{c}{ $K^+$ and $K_1$ exchanges} \\
  \cline{2-3}  \cline{5-7}\\[-2ex]
   &$g_{K^+\Lambda p}$&$\chi^2/N$~~~ &&$g_{K^+\Lambda p}$& $g^V_{K_1^+\Lambda p}$&$\chi^2/N$ \\[1ex] \hline\\[-2ex]
   $0^{\rm th}$ & $-5.09 $ & $1.10$& &$-5.09 $ & $1.49 $& $1.11$\\
   $2^{\rm nd}$ & $-5.32 $ & $0.88$& &$-9.55 $ & $5.93 $& $0.72$ \\
   $4^{\rm th}$ & $-5.32 $ & $0.86$& &$-8.38 $ & $5.42 $& $0.72$\\
   $6^{\rm th}$ & $-5.32 $ & $0.86$& &$-8.89 $ & $6.46 $& $0.72$\\
   $8^{\rm th}$ & $-5.33 $ & $0.86$& &$-9.07 $ & $6.79 $& $0.72$\\
   $10^{\rm th}$& $-5.33 $ & $0.86$& &$-9.13 $ & $6.89 $& $0.72$\\ 
   $12^{\rm th}$& $-5.33 $ & $0.86$& &$-9.15 $ & $6.93 $& $0.72$ \\
   $14^{\rm th}$& $-5.33 $ & $0.86$& &$-9.08 $ & $6.79 $& $0.72$\\
   \hline\hline
  \end{tabular}
\end{table}

\begin{figure}[t]
\centering
\includegraphics[width=10cm]{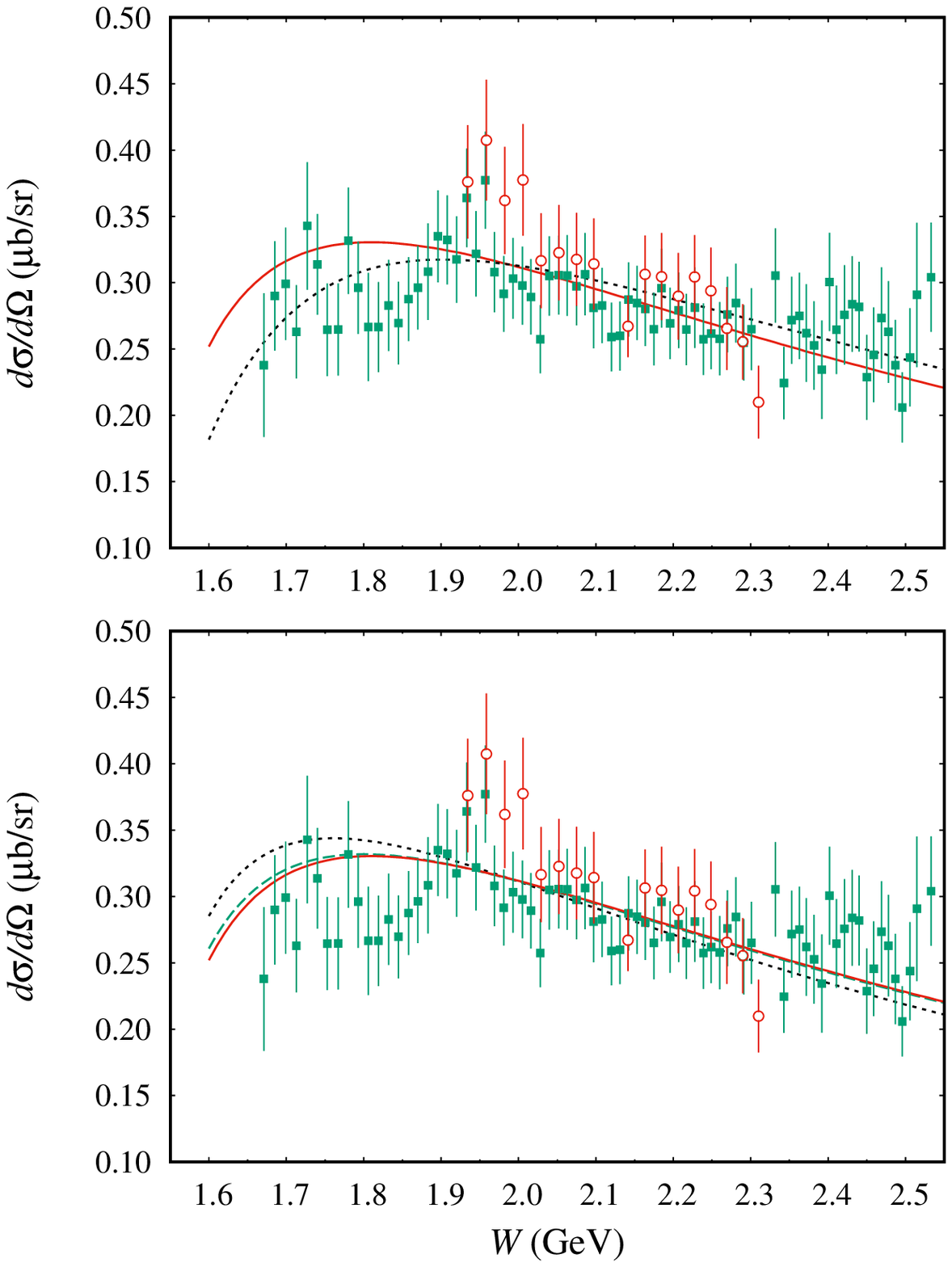}
\caption{{\bf (Top panel)}
  Differential cross section at the forward direction  obtained
  from the $K^+$ intermediate state in $t$-channel (solid line) and 
  from the combination of $K^+$ and $K_1$ intermediate states 
  (dotted line). \\
  {\bf (Bottom panel)}
  Differential cross section at the forward direction  obtained
  from the $0^{\rm th}$ order (dotted line), $2^{\rm nd}$ order 
  (dashed line), and $14^{\rm th}$ order (solid line) expansion. 
  Experimental data are taken from the CLAS collaboration (solid squares)
  \cite{CLAS} and  LEPS collaboration (open circles) \cite{LEPS}.}
\label{Fig:forward}  
\end{figure}

In the present analysis, we only have the coupling $g_{K^+\Lambda p}$
as the unknown parameter, which can be directly extracted by fitting 
the model prediction to the experimental cross section obtained from
the CLAS and LEPS collaborations. To this end the closest available data 
are within $18^\circ<\theta<25^\circ$. The result of extraction 
is listed in Table.\ref{tab:forward_result}, where we can see that
the coupling constant is relatively consistent to all orders. 
As expected, the $\chi^2/N$ decreases as we go to higher orders. 
Actually, the sign of the coupling $g_{K^+\Lambda p}$ cannot be determined
from the fitting process because the amplitude is calculated from
the squared of the coupling. Therefore, we fix the sign by using
the SU(3) prediction \cite{Adelseck:1990ch}, although 
Table \ref{tab:forward_result} exhibits that the extracted value 
is smaller than the SU(3) prediction by almost 50\%. 
Nevertheless, we also note that such values were commonly
obtained in the kaon photoproduction analyses without including hadronic
form factors \cite{Mart:1995wu,adelseck85}.
To remind the reader, within 20\% symmetry breaking
the SU(3) predicts \cite{Adelseck:1990ch}
\begin{eqnarray}
  \label{eq:su(3)}
  -10.6 \leq g_{K^+\Lambda p} \leq -15.6  ~.
\end{eqnarray}
The vector meson coupling $g^V_{K_1^+\Lambda p}$ can be also
estimated by utilizing the SU(3) symmetry. However , since
the extracted coupling from fitting process is the product
$g_{K_1^+ K^+\gamma} \, g^V_{K_1^+\Lambda p}$, 
there is a further uncertainties coming from the determination
of the electromagnetic coupling $g_{K_1^+ K^+\gamma}$.
Thus, we do not pursue to compare our present result with the
SU(3) prediction in this case.

By adding the $K_1(1270)$ exchange to the existing $K^+$ one, 
the considered coupling constants become 
the $g_{K^+\Lambda p }$ and $g^V_{K_1^+\Lambda p}$.
The extracted values along with the corresponding $\chi^2$ are
given in the last three columns of 
Table \ref{tab:forward_result}. It is apparent from this Table that
the $K_1$ exchange is very important in the forward angle kaon 
photoproduction, since by adding the $K_1$ exchange 
the $\chi^2$ decreases and the $g_{K^+\Lambda p }$
increases approaching the SU(3) value. To our knowledge, previous 
calculation has shown that the  $K_1$ exchange is very important
to bring the extracted  $g_{K^+\Lambda p }$ closer to the SU(3)
prediction \cite{Adelseck:1988yv}. Therefore, our present finding 
corroborates this result.

As stated before, the sign of 
coupling constants cannot be determined from the numerical calculation.
Different from the $K_1$ exchange, the $K^*$ intermediate state has 
much smaller contribution in our calculation. The corresponding couplings,
$g^V_{K^* \Lambda N }$ and $g^T_{K^*\Lambda N }$, should have very large 
values in order to produce a small effect in the calculated cross section. 
Thus, we do not show the corresponding values and discuss them here.

Compared to the result obtained from the single $t$-channel intermediate
state, adding the $K_1$ exchange seems to slow down the convergence rate 
as can be seen by comparing the second and fourth columns of
Table \ref{tab:forward_result}. In the single intermediate state
the convergence is already achieved by calculating the amplitude
up to second order. On the contrary, the result obtained from
using both $K^+$ and $K_1^+$ mesons seems to be not convergent
even at the 14$^{\rm th}$ order calculation.

The sensitivity of the $K_1$ exchange is depicted in the top panel
of Fig.~\ref{Fig:forward}. It is apparent from this panel that the
contribution of this state is significant near the threshold region,
where its effect is shifting the calculated cross section closer
to experimental data. The convergence of the squared amplitude expansion 
is displayed in the bottom panel of Fig.~\ref{Fig:forward}. Obviously,
after the second order expansion the expansion is practically convergent.
This is also supported by the fact that the error bars of the present 
data cannot
resolve the the differences produced by different order calculations
(see the bottom panel of Fig.~\ref{Fig:forward}).

To conclude this Section we might safely say that as in the case of the
determination of pion electromagnetic form factor from the $t$-channel pion 
electroproduction \cite{Mart:2008sw}, the extraction of the leading
kaon coupling constant  $g_{K^+\Lambda p}$ can be approximated by only 
using the $t$-channel intermediate states $K$ and $K_1$. 

\section{Kaon Photoproduction at Backward Angles}
\label{sec:backward}

In general, the differential cross section obtained from the $u$-channel 
contribution is relatively smaller than that obtained from the $K^+$ 
exchange in the $t$-channel. In the backward direction, however, 
the magnitude of
Mandelstam variable $|u|$ is very small and, as a consequence, the 
corresponding contribution becomes significant and can be expected to 
dominate all contributions \cite{Morino:2013raa}. Note that 
different from the $t$-channel term, the $u$-channel amplitude 
is individually gauge invariant, since the exchanged particle
$\Lambda$ is neutral and, therefore, the $u$-channel amplitude  
contributes only the magnetic moment term.

\begin{figure}[t]
\centerline{%
\includegraphics[width=10.5cm]{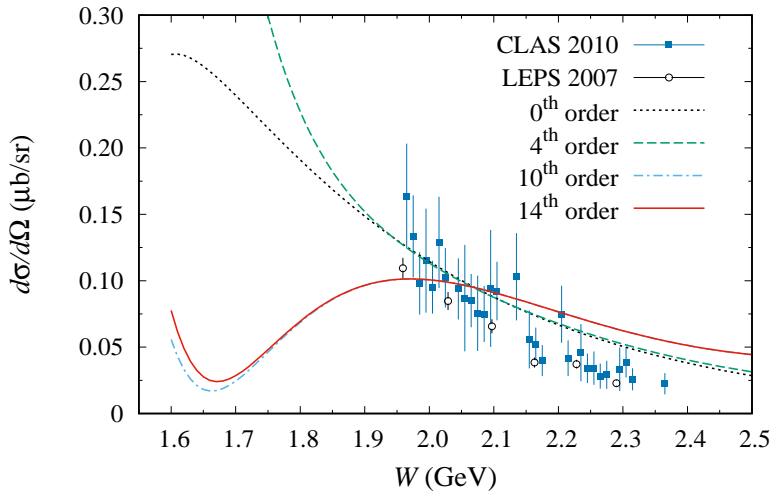}}
\caption{Differential cross section in the backward direction obtained
  from different orders of expansion compared with experimental data. 
  Experimental data are taken from the CLAS  
  (open circles \cite{CLAS2010}) and LEPS (solid squares
  \cite{Hicks:2007zz})  collaborations.} 
\label{Fig:backward}
\end{figure}
\begin{table}[t]
\centering
  \caption{The leading coupling constants in $K^+\Lambda$
    photoproduction, $g_{K^+\Lambda p}$ and  $g_{K^+\Sigma^0 p}$,
    extracted from the backward angle data 
    for different orders of expansion. }
  \label{tab:backward_result}
  \begin{tabular}[c]{cccc}
  \hline\hline\\[-2ex]
   Order   & $g_{K^+\Lambda p}$&  $g_{K^+\Sigma^0 p}$& $\chi^2/N$ \\[0.5ex]
   \hline\\[-2ex]
   $0^{\rm th}$  & \,$-172.9 \pm 62.5$~~  & $-89.7\pm 34.6   $& $26.3$\\
   $2^{\rm nd}$  & ~~~$-1.58 \pm 1340.3 $& ~~~$4.80\pm 697.1   $& $56.6$ \\
   $4^{\rm th}$  & $-457.7 \pm 121.9 $& $-236.8\pm 66.3~~  $& $27.1$\\
   $6^{\rm th}$  & ~~$-694.0 \pm 2160.5$\, & \,$-360.1\pm 1152.2$& $36.8$\\
   $8^{\rm th}$  & $-693.3 \pm 825.9 $& $-360.2\pm 442.9~ $& $33.2$\\
   $10^{\rm th}$ & $-695.6 \pm 929.4 $& $-361.3\pm 498.0~ $& $33.7$\\ 
   $12^{\rm th}$ & $-695.5\pm 926.9  $& $-361.3\pm 496.6~ $& $33.7$ \\
   $14^{\rm th}$ & $-695.5\pm 924.4  $& $-361.3\pm 496.3~ $&$33.7$\\
   \hline\hline
  \end{tabular}
\end{table}

Since we have two exchange particles in the $u$-channel, i.e., $\Lambda$
and $\Sigma^0$,  the unknown parameters in this case are 
$g_{K^{+}\Lambda p}$ and $g_{K^{+}\Sigma^0 p}$ couplings. These 
coupling constants were extracted from fitting the kaon photoproduction 
data with $152^\circ <\theta <161^\circ$.  The result is shown in 
Fig.~\ref{Fig:backward}, where we compare the calculated cross section
obtained from the expanded squared amplitude with 0$^{\rm th}$ up to 
14$^{\rm th}$ orders with the CLAS 2010 and LEPS 2007 data. 
From Fig.~\ref{Fig:backward} we can see that the cross sections 
obtained from expansions with different orders show a large
variance only at low energies, where unfortunately no data are available in
this kinematics. Therefore, experimental measurement to this
end is strongly required in order to check whether or not the 
high order expansions are decisive in the backward direction.

The extracted coupling constants  along with the corresponding 
$\chi^2/N$ obtained from expansions with  different orders 
are listed in Table~\ref{tab:backward_result}. 
In contrast to the case of forward angles, in the backward 
direction both the $\chi^2/N$ and the extracted coupling constants are
very large. The latter are even much larger than the prediction of SU(3).
This result indicates that photoproduction of kaon
at the backward angles cannot be explained by merely using 
the $u$-channel. We observe that the coupling constants can be
greatly decreased if we add the $s$-channel contribution. 
Previous analyses have shown that including hyperon resonances
could also alleviate this problem \cite{Mart:2013hia,Suciawo:2017wtv}.

\section{Summary and Conclusions}
\label{sec:summary}
We have investigated the convergence of the expansion of squared PV 
amplitude for kaon photoproduction at threshold. We found that the 
expanded PV amplitude starts to converge at $9^{\rm th}$ order. Should 
we use solely the zeroth and first order expansions, then the main 
coupling constant $g_{K\Lambda p}$ must be much smaller than the 
SU(3) prediction, if the calculated cross section is fitted to 
experimental data close to threshold. Therefore, our present 
investigation concludes that LET using the expansion of lowest 
order $m_K/m$ would be difficult to derive. We have also extracted 
the leading coupling constants in kaon photoproduction by the expanding 
the squared PV amplitude at forward and backward directions by solely 
utilizing the $t$- and $u$-channel amplitudes, respectively. It is found 
that, unless the $K_1$ vector meson was included, 
the extracted coupling constants in the forward direction are 
smaller than the SU(3) prediction. Adding the $K_1$ significantly
improves the agreement of our calculation with the SU(3) prediction.
In the backward direction we found a different result, 
the extracted coupling constants are much larger than the SU(3) value. 

\section{Acknowledgments}
This work has been supported by the PITTA A Grant, Universitas Indonesia, 
under contract No. NKB-0449/UN2.R3.1/HKP.05.00/2019.

\newpage 
\appendix
\section{ Expansion of the Transition Amplitude}\label{append:Amplitude}
As described in Section \ref{sec:expansion} we expand the square of the 
amplitude according to

$\mathcal{|M|}^2 = \sum_{n=0}^{20} \mathcal{|M|}_{(n)}^2$
where $\mathcal{|M|}_{(n)}^2\propto x^n$.  
To simplify the formulas we define the following quantities.
\begin{eqnarray}
m &=& (m_p+m_\Lambda)/2 ~,\\
x &=& m_K/m ~,\\
y &=& (m-m_\Sigma)/m~\\
\kappa  &=& \kappa_p +\kappa_\Lambda ~,\\
g_\Lambda &=& eg_{K^+\Lambda p} ~,\\
g_\Sigma &=& eg_{K^+ \Sigma^0 p} ~,\\ 
G_\Lambda &=&\kappa g_{\Lambda} ~,\\
G_\Sigma &=& \kappa_T  g_{\Sigma } ~,\\
\label{eq:GKS_coupling}
G_{K^*} &=& (G^T_{K^*}+ G^V_{K^*})/Mm_{K^*}^2 ~,\\
\label{eq:GK1_coupling}
G_{K_1} &=& G^V_{K_1}/Mm_{K_1}^2 ~.
\end{eqnarray}
The $n$-th order of the expanded amplitude is given by
 \begin{eqnarray}
   \label{eq:zeroth-order}
 |\mathcal{M}|^2_{(0)} &=&4  g_{\Lambda}^2 +4 g_\Lambda G_\Sigma y^2~.\\
   \label{eq:first-order}
|\mathcal{M}|^2_{(1)}&=& -(4  g_{\Lambda}^2~+4 g_\Lambda G_\Sigma y- 2g_\Lambda G_\Sigma y^2) x ~.\\
|\mathcal{M}|^2_{(2)}&=&\left[2  g_{\Lambda} (2g_\Lambda+G_\Lambda+ G_\Sigma)+ G_\Sigma y(g_\Lambda  -g_\Lambda y+ G_\Lambda y+2 G_\Sigma y)~~~~\right.\nonumber\\
&&\left.+8G_{K_1}  m^3(2g_{\Lambda}+g_\Sigma y^2)\right] x^2 ~.\\
|\mathcal{M}|^2_{(3)}&=&\left[ \frac{1}{2}  (8 g_\Lambda^2+4 g_\Lambda G_\Lambda +4 g_\Lambda  G_\Sigma +g_\Lambda  G_\Sigma y+2 G_\Lambda  G_\Sigma  y+2  G_\Sigma^2  y\right.\nonumber\\
&&\left.-2 g_\Lambda G_\Sigma  y^2-g_\Lambda  G_\Sigma  y^2-2  G_\Sigma^2 y^2)- G_{K^*} m^3 (2  g_\Lambda + G_\Sigma y^2)\right.\nonumber\\
&&\left.+4G_{K_1}  m^3 (4  g_\Lambda +2 G_\Sigma   y- G_\Sigma   y^2)\right]x^3 ~.
\end{eqnarray}
\begin{eqnarray}
|\mathcal{M}|^2_{(4)}&=&-\left[-\frac{1}{16} (4 g_\Lambda^2 (4+\kappa)^2+G_\Sigma (4+4 y-7 y^2)+2 
G_\Sigma (16g_\Lambda+2g_\Lambda y\right.\nonumber\\
&&\left.-7g_\Lambda  y^2+G_\Lambda (4+2 y-2 y^2)))+
4G_{K^*} m^3 (3  g_\Lambda +G_\Sigma y  )\right.\nonumber\\
&&\left.-2 G_{K_1}  m^3(10  g_\Lambda +2  G_\Lambda +2G_\Sigma +G_\Sigma  y)-16
G_{K_1}^2 m^6\right.\nonumber\\
&&\left.+\frac{8 G_{K_1}  m^5 (2  g_\Lambda+G_\Sigma y^2)}{m_{K_1}^2})\right]x^4 ~.
 \end{eqnarray}
\begin{eqnarray}
 |\mathcal{M}|^2_{(5)}&=&-\left[\frac{1}{8} (2 g_\Lambda^2 (4+\kappa)^2+G_\Sigma^2 (2+y+53 y^2)+ G_\Sigma (16g_\Lambda+g_\Lambda y+2g_\Lambda y^2\right.\nonumber\\
 &&\left.+G_\Lambda (4+y+26 y^2))+16
 G_{K_1}^2 m^6-G_{K^*}
   m^3  (2 g_\Lambda (8+\kappa)+G_\Sigma (2\right.\nonumber\\
 &&\left.+3 y)) + G_{K_1} m^3 (4 g_\Lambda (6+\kappa)+G_\Sigma (4+3
 y- y^2))- 16  G_{K^1}G_{K^*} m^6\right.\nonumber\\
 &&\left.-
  \frac{4 G_{K_1} m^5(8  g_\Lambda +2 G_\Sigma y+G_\Sigma  y^2)}{m_{K_1}^2}+
 \frac{4 G_{K^*} m^5(2  g_\Lambda +G_\Sigma y^2)}{ m_{K*}^2}\right]x^5 ~.\nonumber\\
\end{eqnarray}
\begin{eqnarray}
 |\mathcal{M}|^2_{(6)}&=&-\left[-\frac{1}{64} (16 g_\Lambda^2 (4+\kappa)^2+G_\Sigma^2 (16+4 y-63 y^2)+2 
 G_\Sigma (64g_\Lambda\right.\nonumber\\
 &&\left.+2g_\Lambda y-g_\Lambda y^2+G_\Lambda (16+2 y-15 y^2)))-
 \frac{1}{16}  G_\Sigma (-10 g_\Lambda+4 G_\Lambda\right.\nonumber\\
 &&\left. +11 G_\Sigma) y^2-16 G_{K^*}^2 m^6 -24 G_{K_1}^2 m^6+ 24 G_{K^*} G_{K_1} m^6\right.\nonumber\\
  &&\left.- 8G_\Sigma G_{K_1} m^3  y^2-\frac{8G_{K_1} m^7
 (2  g_\Lambda -4 G_{K_1} m+G_\Sigma y^2)}{m_{K_1}^4}\right.\nonumber\\
 &&\left.-\frac{4 G_{K^*} m^5(5  g_\Lambda +G_\Sigma  y+G_\Sigma  y^2)}{m_{K*}^2}-\frac{1}{4}(G_{K_1} m^3 (4 g_\Lambda (28+5 \kappa)\right.\nonumber\\
 &&\left.+G_\Sigma (20+12
   y-37 y^2)))+
  \frac{1}{2 }( G_{K^*} m^3 (g_\Lambda (40+6 \kappa)+G_\Sigma
    (6+6 y\right.\nonumber\\
  &&\left.-y^2)))+\frac{2 G_{K_1} m^5 ( 2 g_\Lambda (13+\kappa)+G_\Sigma (2+5 y+2 y^2))}{ m_{K_1}^2}\right]x^6 ~.\nonumber\\
 \end{eqnarray}
 \begin{eqnarray}
 |\mathcal{M}|^2_{(7)}&=&- \left[\frac{1}{64} (16 g_\Lambda^2 (4+\kappa)^2+G_\Sigma^2 (16+2 y-15 y^2)+2 
 G_\Sigma (64g_\Lambda+g_\Lambda y\right.\nonumber\\
  &&\left.-19g_\Lambda y^2+G_\Lambda (16+y-6 y^2)))-\frac{4 G_{K^*} m^7 (2 g_\Lambda+G_\Sigma y^2)}{m_{K*}^4}\right.\nonumber\\
  &&\left.-
   \frac{64G_{K_1}^2
     m^8 }{m_{K_1}^2}+
 \frac{4G_{K_1}
   m^7 (12 e g_\Lambda +2 G_\Sigma  y+3G_\Sigma  y^2)}{m_{K_1}^4}\right.\nonumber\\
 &&\left.+16 G_{K_1}^2 m^6+\frac{G_{K_1} m^5 (-4 g_\Lambda (3+2 \kappa)+G_\Sigma (8-11 y+7 y^2))}{m_{K_1}^2}\right.\nonumber\\
 &&\left.+
 \frac{1}{2}(G_{K_1} m^3 (4 G_\Lambda +G_\Sigma (4-4 y+3 y^2)))-36
      G_{K_1} G_{K^*} m^6\right.\nonumber\\
  &&\left.+8 G_{K^*}^2 m^6+\frac{G_{K^*}
  m^5 (2 g_\Lambda (18+\kappa)+G_\Sigma (2+7 y+4 y^2))}{m_{K*}^2}\right.\nonumber\\
    &&\left.-
 \frac{1}{8}(G_{K^*} m^3 (32 g_\Lambda (6+\kappa)+G_\Sigma
  (32+24 y-7 y^2)))+16 G_{K_1}^2 m^6\right.\nonumber\\
    &&\left.{G_{K_1} m^3 (4 g_\Lambda (8+\kappa)+G_\Sigma (4+5 y-3 y^2))}{m_{K_1}^2}+\frac{16
  G_{K^*} G_{K_1} m^8}{m_{K_1}^2 }\right.\nonumber\\
   &&\left.+\frac{16 G_{K^*} G_{K_1} m^8}{m_{K^*}^2}-
 \frac{4 G_{K_1} m^5 (16  g_\Lambda+6  G_\Sigma y- G_\Sigma y^2)}{m_{K_1}^2}\right]x^7 ~.\nonumber\\
\end{eqnarray}
\begin{eqnarray}
|\mathcal{M}|^2_{(8)}&=&-\left[-\frac{1}{256}(64 g_\Lambda^2 (4+\kappa)^2+G_\Sigma^2 (64+4 y-49 y^2)+2 
G_\Sigma (256g_\Lambda\right.\nonumber\\
&&\left.+2g_\Lambda y-71g_\Lambda y^2+G_\Lambda (64+2 y-21 y^2)))+50 G_{K_1} G_{K^*} m^6 \right.\nonumber\\
&&\left.+
\frac{112 G_{K_1}^2 m^8 }{m_{K_1}^2}-
\frac{2 G_{K_1} m^7 (2 g_\Lambda (25+\kappa)+G_\Sigma (2+9 y+8 y^2))}{m_{K_1}^4}\right.\nonumber\\
&&\left.-\frac{48G_{K_1}^2  m^{10}}{m_{K_1}^4}+\frac{8G_{K_1}  m^9
(2  g_\Lambda+G_\Sigma  y^2)}{m_{K_1}^8}-40 G_{K_1} G_{K^*}
m^8 \right.\nonumber\\
&&\left.+
\frac{8 G_{K^*}^2 m^8 }{m_{K*}^2}+
\frac{4 G_{K^*} m^7 (7 g_\Lambda+G_\Sigma y (1+2 y))}{m_{K*}^4}-\frac{40 G_{K_1} G_{K^*} m^8}{m_{K*}^2}\right.\nonumber\\
&&\left.-
13G_{K^*}^2  m^6 -
\frac{G_{K^*}  m^5 (+2 g_\Lambda (56+5 \kappa)+G_\Sigma (10+20 y+7 y^2))}{2m_{K*}^2}\right.\nonumber\\
&&\left.+
\frac{1}{16 }G_{K^*} m^3 (16 g_\Lambda (28+5 \kappa)+G_\Sigma
(80+48 y-19 y^2))\right.\nonumber\\
&&\left.-164 G_{K_1}^2 m^6+\frac{G_{K_1} m^5 (52 g_\Lambda (8+\kappa)+G_\Sigma (52+64 y+7 y^2))}{4 m_{K_1}^2}\right.\nonumber\\
&&\left.-
\frac{1}{4}G_{K_1} m^3 (4 g_\Lambda (36+7 \kappa)+G_\Sigma (28+12
y-7 y^2))\right]x^8 ~.
\end{eqnarray}
\begin{eqnarray}
|\mathcal{M}|^2_{(9)}&=&-\left[\frac{1}{128} 2 (32 g_\Lambda^2 (4+\kappa)^2+G_\Sigma^2 (32+y-21 y^2)+
G_\Sigma (256g_\Lambda\right.\nonumber\\
&&\left.+g_\Lambda y-68g_\Lambda y^2+G_\Lambda(64+y-19 y^2)))\right.\nonumber\\
&&\left.+
\frac{144 G_{K_1}^2 m^{10}}{m_{K_1}^4}-
\frac{4 G_{K_1} m^9(16  g_\Lambda+2 G_\Sigma G_{K_1}  m^9 y+5  G_\Sigma y^2)}{m_{K_1}^6}\right.\nonumber\\
 &&\left.-\frac{48G_{K_1}^2 m^8 }{m_{K_1}^2}+\frac{G_{K_1} m^7 (4 g_\Lambda (16+3 \kappa)+G_\Sigma (12+7 y-y^2))}{m_{K_1}^4}\right.\nonumber\\
  &&\left.+
 3G_{K_1}^2 m^6 +
\frac{3G_{K_1} m^5 (-4 g_\Lambda (4+\kappa)+G_\Sigma
(-4+y^2))}{4m_{K_1}^2}\right.\nonumber\\
 &&\left.+\frac{1}{32} g_\Lambda G_{K_1} (-1280+8007 \kappa )m^3 y^2+
 \frac{4G_{K^*} m^9 (2  g_\Lambda + G_\Sigma y^2)}{m_{K*}^6}\right.\nonumber\\
 &&\left.-\frac{16 G_{K^*}  G_{K_1}
 m^{10}}{ m_{K*}^4}+\frac{76G_{K_1} G_{K^*} m^8 }{m_{K*}^2}-\frac{16
 G_{K_1} G_{K^*} m^{10} }{m_{K_1}^4}\right.\nonumber\\
  &&\left.+152 G_{K^*}^2 m^6+
\frac{G_{K^*} m^5 (8 g_\Lambda (80+9 \kappa)+G_\Sigma (72+104 y+21 y^2))}{8 m_{K*}^2}\right.\nonumber\\
 &&\left.-
 \frac{24 G_{K^*}^2 m^8 }{m_{K*}^2}-
\frac{G_{K^*} m^7 (2 g_\Lambda (32+\kappa)+G_\Sigma
 (2+11 y+12 y^2))}{m_{K*}^4}\right.\nonumber\\
  &&\left.
\frac{1}{32 } G_{K^*} m^3 (64 g_\Lambda (16+3 \kappa)+G_\Sigma
(192+96 y-47 y^2))\right.\nonumber\\
 &&\left.+48 G_{K_1}^2 m^6+\frac{G_{K_1} m^5 (-4 g_\Lambda (31+4 \kappa)-G_\Sigma (16+19 y+y^2))}{m_{K_1}^2}\right.\nonumber\\
 &&\left.+
G_{K_1} G_\Sigma  m^3  (8+3 y-2 y^2)-\frac{16G_{K_1}
 G_{K^*} m^{10} }{m_{K_1}^2m_{K*}^2}\right.\nonumber\\
 &&\left.+
\frac{1}{32}G_{K_1} m^3 (1280 g_\Lambda
(1+y^2)+G_\Lambda (256-8007 y^2))\right.\nonumber\\
 &&\left.-
 \frac{128 G_{K_1}^2 m^8 }{m_{K_1}^2}+
\frac{4 G_{K_1} m^7 (28 e g_\Lambda+6 e G_\Sigma y+5 e G_\Sigma y^2)}{m_{K_1}^4}\right.\nonumber\\
 &&\left.+
\frac{76 G_{K_1} G_{K^*} m^8 }{m_{K_1}^2}-66 G_{K_1} G_{K^*} m^6 \right]x^9 ~.
\end{eqnarray}

\end{document}